# A New PID Neural Network Controller Design for Nonlinear Processes

Ali Zribi
Assistant professor
ali_zribi@yahoo.fr

Mohamed Chtourou
Professor
Mohamed.Chtourou@enis.rnu.tn

Mohamed Djemel
Professor
Mohamed.Djemel@enis.rnu.tn

**Abstract:**

In this paper, a novel adaptive tuning method of PID neural network (PIDNN) controller for nonlinear process is proposed. The method utilizes an improved gradient descent method to adjust PIDNN parameters where the margin stability will be employed to get high tracking performance and robustness with regard to external load disturbance and parameter variation. Simulation results show the effectiveness of the proposed algorithm compared with other well-known learning methods.

1.  **Introduction:**

Recently, PIDNN controller is one of the popular methods used for control complexes systems. Several robust and auto tuning techniques have been proposed in order to further improve the control and robust performance of the PIDNN controller [1,2,3,4]. In [5], an adaptive PIDNN controller was presented. To improve the convergent speed and to prevent the weights getting trapped into local optima, the particle swarm optimization (PSO) algorithm was adopted to initialize the neural network. Empirical results illustrate the effectiveness of the proposed algorithm. But, during the weight initialization, PSO algorithm takes a long time. Cong et Liang [6] studied the effect of learning rates, of the gradient based methods, to stability of closed-loop system is analyzed. The weights of the NNPID are updating according to errors caused by uncertain factors of the controlled system. The robustness of the implemented controllers is investigated under parameter uncertainty however an improving in the convergence speed should be established. In [7], Ho et al. formulated an optimization problem of establishing a fuzzy neural network model (FNNM) for efficiently tuning PID controllers of a plant with under-damped responses. However, the proposed algorithm requires significant effort due to the number of parameters to be determined prior to the implementation. Vikas et al. [8] proposed a neural network based PID like controller composed of a mixed locally recurrent neural network and contains at most three hidden nodes which form a PID like structure. The presented controller is easy to implement. But, the main drawback in such strategy is in number of parameters to be determined prior to the training which requires a priori process knowledge.

In this paper, the key ideas explored are the use of the gradient descent method and the stability margin for training the weights of the PIDNN to make the learning algorithm converges much faster and to increase the robustness of the control algorithm than other well-known algorithms. It consists in introducing of an improved momentum which helps to have good robustness with regard to external load disturbance and parameters variation. The contribution of the presented momentum is depended on the stability margin measure. It accelerates learning mainly when the global stability of the controlled system is not guaranteed and as well as when the desired tracking performance is not achieved.

This paper is organized as follows: In Section 2, related background is shortly reviewed. Section 3 explains the control system architecture and the proposed tuning method for the PIDNN. To check the ability of the proposed approach, two examples have been considered in Section 4. Finally, Section 5 provides the conclusion.

## 2. Review of PID controller and margin stability:

### 2.1. PID

The strategy of *PID* control has been one of sophisticated and most frequently used methods in the industry. By taking the time-derivative and discretizing the resulting equation and of the both sides of the continuous-time *PID* equation, a typical discrete-time *PID* controller can be expressed as:

$$u(k) = u(k-1) + K_p \ (e(k)-e(k-1)) + \frac{K_I T}{2} \ (e(k)+e(k-1)) + \frac{K_D}{2}(e(k)-2e(k-1)+e(k-2)) \quad (1)$$

where *u(k)* is the control effort at time *k*, $K_P$, $K_I$ and $K_D$ are proportional, integral and derivative gains, respectively. *T* is the sampling period. *e(k)* is the tracking error defined as $e(k)=y_d(k)-y(k)$ ; $y_d(k)$ is the desired plant output, *y(k)* is the actual plant output. The *PID* controller Eq. (1) can also be expressed in the following form [9,10,11,12].

$$\Delta u(k) = u(k) - u(k-1) = K_p \ e_p(k) + K_I \ e_I(k) + K_D \ e_D(k) \quad (2)$$

Where

$$\begin{aligned} e_p(k) &= e(k) - e(k-1) \\ e_I(k) &= \frac{T}{2}\left(e(k) + e(k-1)\right) \\ e_D(k) &= \frac{1}{T}\left(e(k) - 2\,e(k-1) + e(k-2)\right) \end{aligned} \quad (3)$$

### 2.2. Margin stability:

Let *P* a linear system with a normalized right coprime factorization $P = NM^{-1}$ presented as a left coprime factor perturbed system described as follows:

$$P_\Delta = \left(M + \Delta M^{-1}\right)\left(N + \Delta N^{-1}\right) \quad (4)$$

Where uncertainty $\Delta M, \Delta N \in RH_\infty$ and $\left\| \begin{bmatrix} \Delta M \\ \Delta N \end{bmatrix} \right\|_\infty \leq \varepsilon$.

It has been shown in [13,14] that the system is robustly stable if:

$$\left\|\begin{bmatrix} I \\ C \end{bmatrix}[I+PC]^{-1}M^{-1}\right\| = \left\|\begin{bmatrix} I \\ C \end{bmatrix}[I+PC]^{-1}\begin{bmatrix} I & P \end{bmatrix}\right\| \leq \frac{1}{\varepsilon} \qquad (5)$$

Where *C* be a stabilizing controller of *P*.

Thus, we can define the stability margin of the system as follows:

$$b_{P,C} = \left\|\begin{bmatrix} I \\ C \end{bmatrix}[I+PC]^{-1}\begin{bmatrix} I & P \end{bmatrix}\right\| \qquad (6)$$

Where $0 \leq b_{P,C} \leq 1$ and is used as an indication of robustness to unstructured perturbations. A controller with a value of greater than *0.3* generally indicate good robustness margins [15].

### 3. Learning algorithm

#### *3.1. Control system architecture*

The structure of the newly proposed control algorithm of nonlinear *PID*-based neural networks is shown in Fig. 1. The *PID* is placed in cascade with the plant neural network model. The error signal obtained as the difference between the plant model and desired plant output is used to adjust only the *PID* parameters. In the proposed adaptive control scheme, the model plant weights are adjusted of-line then the fixed weights will be used only to adjust on-line the gains of the *PID* controller. The presented control algorithm has simple structure and provides a gain in computation time.

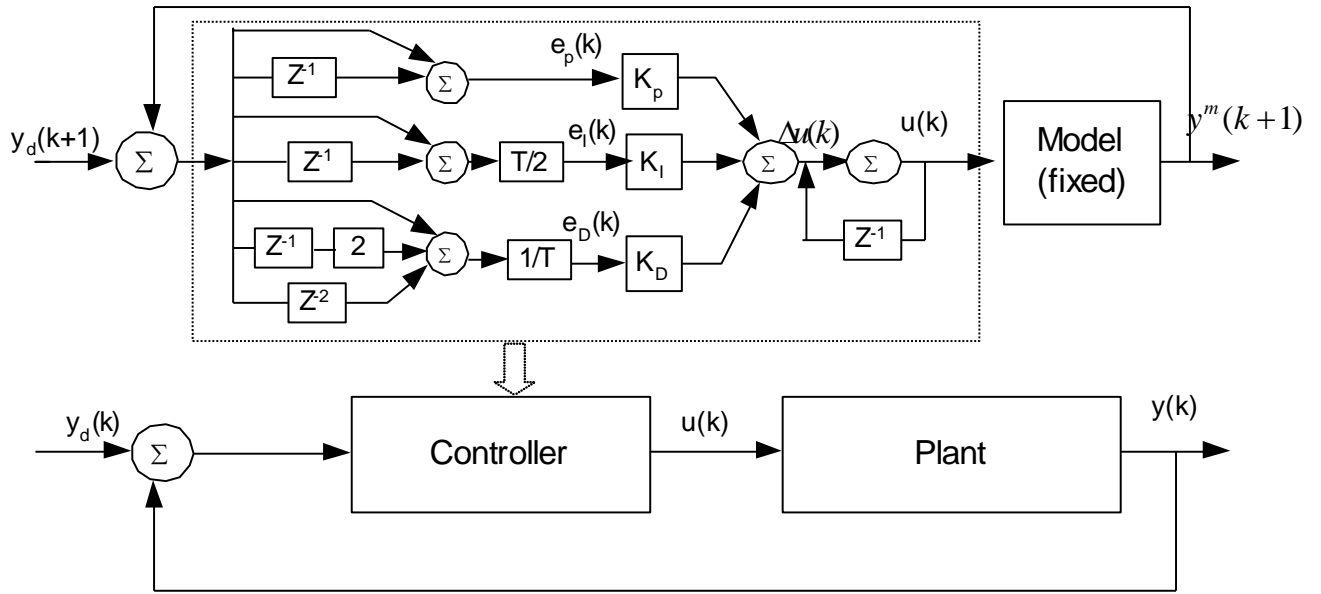

**Fig. 1.** Proposed *PID* control system architecture.

To get the neural network model plant, a feedforward neural network is used to learn the system and back-propagation algorithm is employed to train the weights. The block diagram of identification system is shown in Fig. 2 [16].

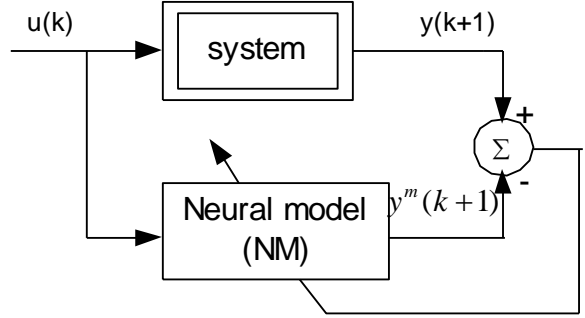

**Fig. 2.** The on-line Neural model training.

Assume that the unknown nonlinear system to be considered is expressed by:

$$y(k+1) = f\left[y(k), y(k-1),\ldots, y(k-n+1), u(k), u(k-1),\ldots, u(k-m+1)\right] \quad (7)$$

where *y(k)* is the scalar output of the system, *u(k)* is the scalar input to the system, *n* and *m* are the output and the input orders respectively, f[...] is the unknown nonlinear function to be estimated by a Neural Network.

The neural model for the unknown system can be expressed as:

$$y^m(k+1) = \hat{f}\left[y(k), y(k-1),\ldots, y(k-n+1), u(k), u(k-1),\ldots, u(k-m+1)\right] \quad (8)$$

Where $y^m$ is the output of the neural model and $\hat{f}$ is the estimate of *f*.

The weights of the neural model are adjusted to minimize the cost function given by:

$$E = \frac{1}{2}\left(y(k+1) - y^m(k+1)\right)^2 \quad (9)$$

### 3.2. *Adaptation of the PID controller gains*

At each control cycle, the controller gains are adjusted. The cost function and the NM whose weights are now fixed are used to adjust the controller gains. In order to make the learning algorithm converges much faster while guaranteeing stability of the closed-loop system, the stability properties has been investigated in weight update. The idea consists in incorporating in the present weight update some influence of past iterations by using a momentum. The influence of the momentum depends on the stability properties of the closed-loop system. Thus, the influence can became large in cases where a desired stability performances is not well attained (small margin stability).
The weight tuning update for the gains using descent method is given by:

$$\Delta k_{p,I,D}(t) = -\alpha \frac{\partial E(t)}{\partial k_{p,I,D}} + \beta(t)\Delta k_{p,I,D}(t-1) \tag{10}$$

α is the learning rate and β is the momentum rate tuned according the flowing equation:

$$\beta(t) = \beta_0 \exp(-b_{P,C}) \tag{11}$$

Where $\beta_0$ is its initial value.

The presented adaptive momentum tuning method is developed to accelerate the adaptation of the controller parameters when the controller has poor robustness performance and to slow down the adaptation when the controller has good robustness performance.

The derivates in equation (10) is computed as:

$$\begin{aligned}
\frac{\partial E}{\partial k_p} &= \frac{\partial E(t)}{\partial y^m}\frac{\partial y^m}{\partial u(t)}\frac{\partial u(t)}{\partial K_p} = -(y-y^m)\frac{\partial y^m}{\partial u(t)}e_p(t) \\
\frac{\partial E}{\partial k_I} &= \frac{\partial E}{\partial y^m}\frac{\partial y^m}{\partial u(t)}\frac{\partial u(t)}{\partial K_I} = -(y-y^m)\frac{\partial y^m}{\partial u(t)}e_I(t) \\
\frac{\partial E}{\partial k_D} &= \frac{\partial E(t)}{\partial y^m}\frac{\partial y^m}{\partial u(t)}\frac{\partial u(t)}{\partial K_D} = -(y-y^m)\frac{\partial y^m}{\partial u(t)}e_D(t)
\end{aligned} \tag{12}$$

As the hidden and output neuron functions were defined by the logistic sigmoid function $f(x) = 1/(1+\exp(-x))$ then $\frac{\partial y^m}{\partial u(t)}$ is expressed as:

$$\frac{\partial y^m}{\partial u} = y^m(1-y^m)\sum_j W_j^m O_j^m (1-O_j^m) W_{1j}^m \tag{13}$$

$O_j^m$ is the output of $j^{th}$ neuron in the hidden layer of the *NM*, $W_{1j}^m$ and $W_j^m$ are the weights of the *NM* from the input neurons to intermediate layer, and from the intermediate layer to the output.

## 4. Simulation results

### 4.1. *Example 1: isothermal CSTR*

The above algorithm is applied to a simple nonlinear process. The process chosen is an isothermal CSTR in which a first-order irreversible reaction takes place [17,18]. The relevant mass balance is:

$$\frac{dC_A}{dt} = -K\,C_A + (C_{Ai} - C_A)\,u \tag{14}$$

where $C_A$ is the concentration of the reactant and $u$ is the dilution rate. The rate constant $K$ is $0.028\ min^{-1}$ and the initial concentration $C_{Ai}$ is *1.0 mol/L*. The control objective is to regulate $C_A$ at a particular setpoint by manipulating *u*.

We first construct the NM through the task of designing our adaptive *PID* control system. It has two inputs *[$C_A(k)$,$u(k)$]*, six hidden neurons and one output *$C_A(k+1)$*. The initial weight values of the conventional *BP* are set to small random values and a learning rate of *0.1* is used in *BP* algorithm. The obtained *NM*, whose weights are now fixed, will be used to train the *PIDNN*. The initial gains values of the *PIDNN* are as *($K_p$,$K_I$,$K_D$)=(0.5,0.2,1)*, and parameters for the proposed algorithm are *α=0.2, $β_0$=0.8*.

The mean square error (MSE), by the following equation, is used to evaluate the modelling and control performance:

$$E = \frac{1}{2}\sum_{k=1}^{N}\left(y_d(k) - y(k)\right)^2 \tag{15}$$

*N* denotes the control horizon.

Fig. 3 shows the performance of the PIDNN controllers in cases without a momentum term (*MSE=1.21*), with a fixed momentum term (*MSE=1.06*) and with a variable momentum (*MSE=0.95*) adjusted dynamically by the margin stability. From simulation results, the effect of the proposed method is clear where significant improvement can be seen in terms of tracking performance and by evaluating the control performance using the mean square error where the minimum is attained.

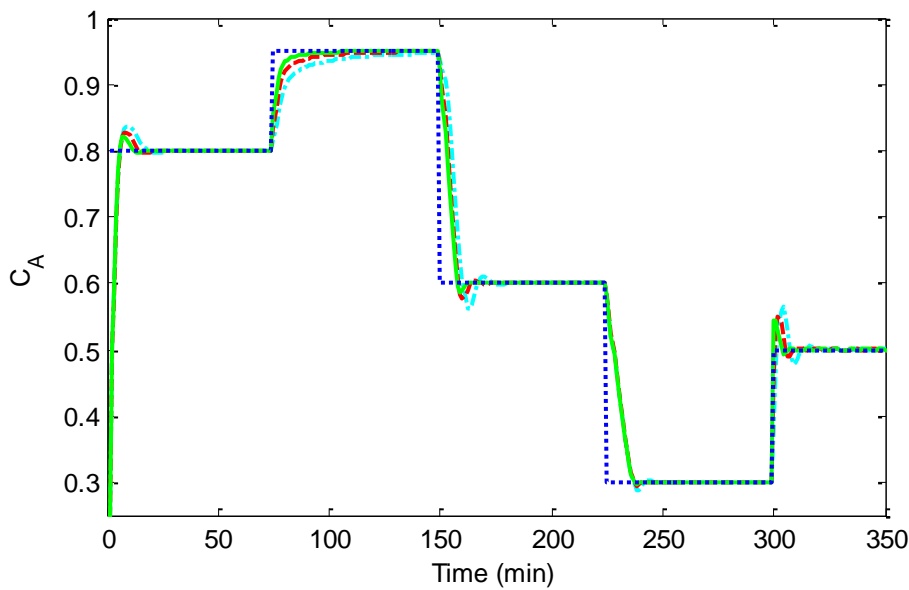

**Fig. 3.** Closed-loop responses and of CSTR using PIDNN:
((dash-dotted) without momentum ;(dashed) with fixed momentum ; (solid) with variable momentum; (dotted) setpoint).

The stability margin for the controlled system using three methods is shown in Fig. 4. It is clear that the proposed controller is better than the other control methods in terms of robust performance. In this regard, the proposed method provides a better tradeoff between the robustness and the tracking performances.

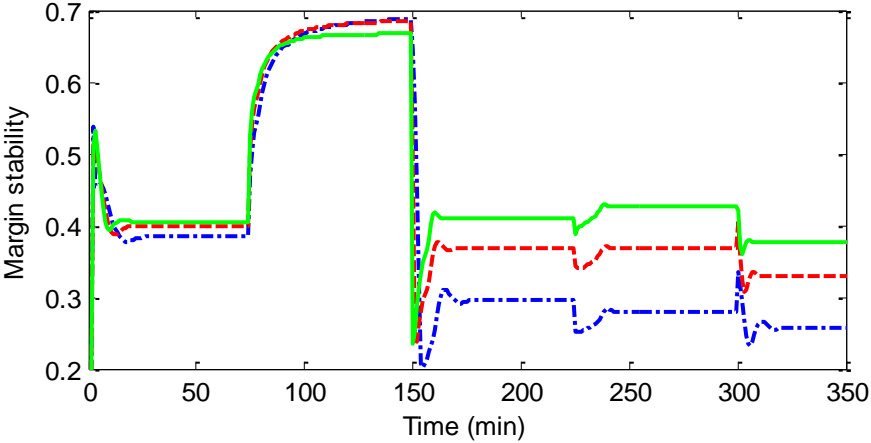

**Fig. 4.** Calculated margin stability ((dash-dotted) without momentum ;(dashed) with fixed momentum ; (solid) with variable momentum).

Fig. 5 shows the disturbance rejection responses of the CSTR for an additive disturbance added at time t=25min. As is seen, our proposed controller performs satisfactorily for disturbance rejection; it can reject disturbances quickly and bring the output back to reference signal precisely.

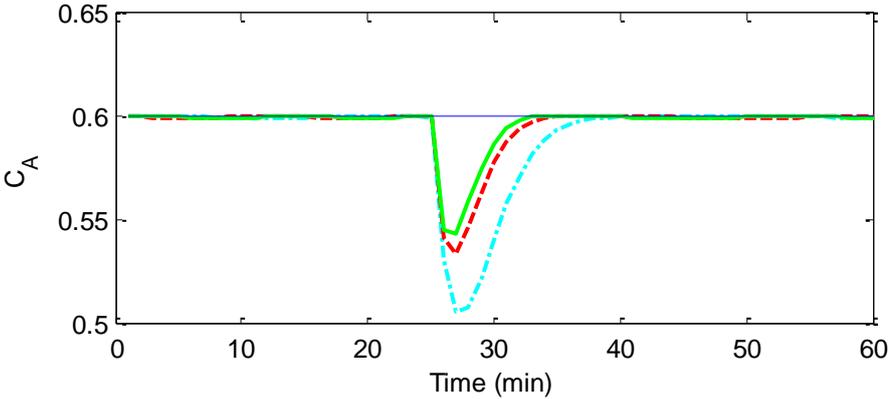

**Fig. 5.** Disturbance rejection responses ((dash-dotted) without momentum ;(dashed) with fixed momentum ; (solid) with variable momentum).

### 4.2. Example 2: Nonisothermal CSTR

Consider a nonisothermal CSTR where an irreversible, first-order reaction takes place [19]. The mathematical model is:

$$\begin{cases} \dot{x}_1 = -x_1 + D_a(1-x_1)\exp\left(\dfrac{x_2}{1+x_2/\gamma}\right) \\ \dot{x}_2 = -x_2 + BD_a(1-x_1)\exp\left(\dfrac{x_2}{1+x_2/\gamma}\right) + \beta(u-x_2) \\ y = x_2 \end{cases} \qquad (16)$$

where $x_1$, $x_2$, and $u$ are the dimensionless reagent conversion, the temperature (output), and the coolant temperature (input), respectively and the nominal values for the constants in are $D_a=0.72$, $B=8$, $\gamma=20$ and $\beta=0.3$.

In the first step, the NM for the presented system has to be constructed. The network structure used has three inputs $x_1(k)$, $x_2(k)$ and $u(k)$, ten hidden neurons and a single output $x_2(k+1)$. The system identification was initially performed with the plant input with amplitude uniformly distributed over the interval *[−2,+2]* and the desired output was normalized in the interval *[0.05, 0.95]*. The obtained *NM* will now be used to design the *PIDNN* controller. The *PIDNN* is trained on-line to minimize the cost function and to adapt to the variations in plant parameters. The initial gains values of the *PIDNN* are as $(K_p, K_I, K_D)=(7.5, 2.5, 1)$, and parameters for the proposed algorithm are: $\alpha=0.1$, $\beta_0=0.9$.

The closed loop responses are shown in Fig. 6 for PIDNN controllers using an adjusted momentum, a fixed momentum or without momentum.

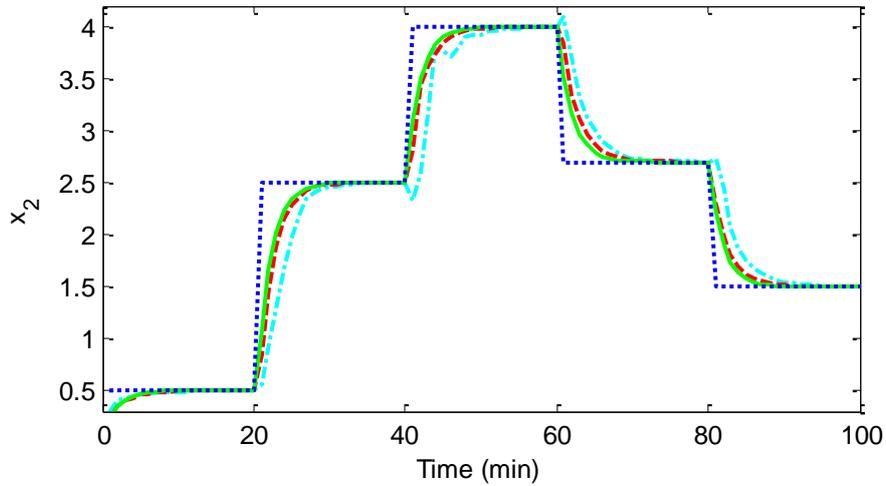

**Fig. 6.** Closed-loop responses and of nonisothermal CSTR using PIDNN: ((dash-dotted) without momentum ;(dashed) with fixed momentum ; (solid) with variable momentum; (dotted) setpoint).

For evaluating the tracking performance of the three controllers the mean square error is used to make the comparisons. Table 1 demonstrates that the proposed controller has a better tracking performance. A decrease in tracking performances is detected using a *PIDNN* with a fixed momentum and a quite bad performance is obtained by a *PIDNN* without a momentum.

**Table 1** MSE results

| PIDNN | Without momentum | With a fixed momentum | With an adjusted momentum |
|---|---|---|---|
| MSE | 69.1 | 60.6 | 50.59 |

Our proposed controller has consistently the best in terms of robust control performance in the whole operating space. A significant decrease in robustness properties is obtained using a *PIDNN* with a fixed momentum. And without a momentum, the *PIDNN* gives poor robust performance.

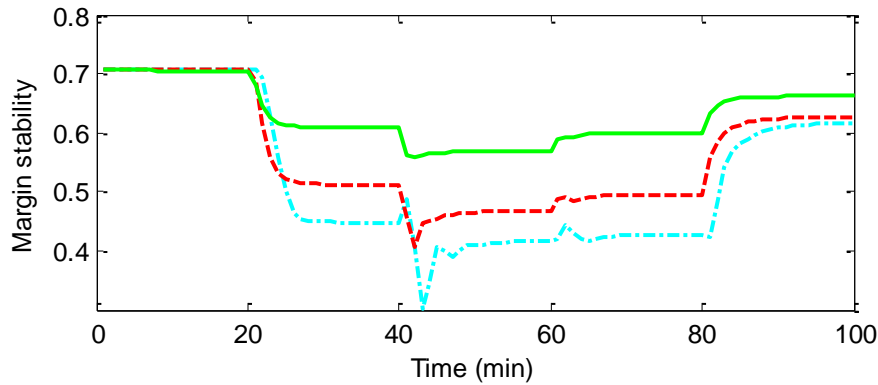

**Fig. 7.** Calculated margin stability ((dash-dotted) without momentum ;(dashed) with fixed momentum ; (solid) with variable momentum)

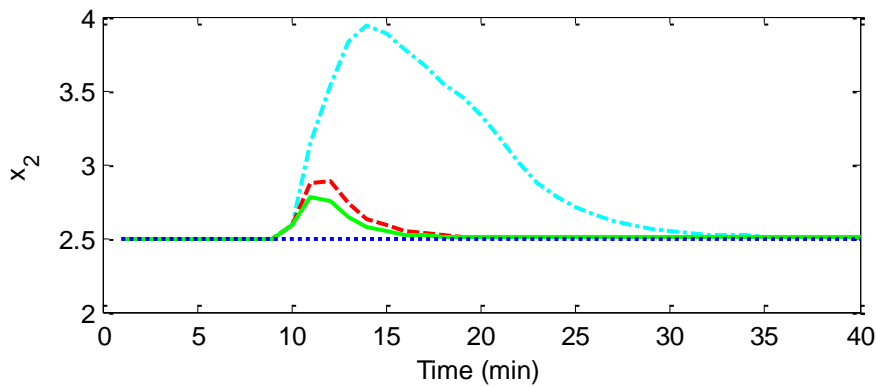

**Figure 8.** Disturbance rejection responses ((dash-dotted) without momentum ;(dashed) with fixed momentum ; (solid) with variable momentum)

The performance of the proposed control system is tested by altering the CSTR parameters $D_a=0.072$ to $D_a=0.09$ at the time index of *8*. As can be seen from Fig. 8, the proposed adaptive controller has the best performance in disturbance rejection. The controller responses of a *PID* without a momentum are sluggish with considerable overshoot which may not be acceptable.

## 5.    Conclusion

In this paper, a novel adaptive tuning method for PIDNN controller is developed to control nonlinear processes in which stability margin is used adjust the PID parameters. Simulation results show that the method presented is not only effective for the set point tracking and the stability of control system but also effective for the process robustness. Since the robustness can be improved by tuning the PID parameters.